# Hyperfine coupling constants of the cesium $7D_{5/2}$ state measured up to the octupole term

**BUBAI RAHAMAN**[1] **AND SOURAV DUTTA**[1,*]

[1]*Tata Institute of Fundamental Research, 1 Homi Bhabha Road, Colaba, Mumbai 400005, India*
*Corresponding author: sourav.dutta@tifr.res.in*

**We report the measurement of the hyperfine splitting in the $7D_{5/2}$ state of $^{133}$Cs using high resolution Doppler-free two-photon spectroscopy enabled by precise frequency scans using an acousto-optic modulator. All the six hyperfine levels are resolved in our spectra. We determine the hyperfine coupling constants $A$ = -1.70867(62) MHz and $B$ = 0.050(14) MHz which represent over 20-times improvement in the precision of both $A$ and $B$. Moreover, our measurement is sufficiently precise to put bounds on the value of the magnetic octupole coupling constant $C$ = 0.4(1.4) kHz for the $7D_{5/2}$ state. We additionally report the measurement of ac Stark shift [-46±4 Hz/(W/cm$^2$)], collisional shift and pressure broadening which are important for optical frequency standards based on the $6S_{1/2} \to 7D_{5/2}$ two-photon transition.**

https://doi.org/10.1364/OL.469086

The hyperfine splitting (HFS) arises from the electron-nucleus interaction and therefore the measurement of HFS provides information about the nuclear structure as well as the electronic wave function in the vicinity of the nucleus [1,2]. Comparison of the experimentally measured HFS with those calculated theoretically allows benchmarking of theoretical models; and deviations often indicate the presence of new interactions which are subsequently included in refinement of the model [3–5]. The cesium atom is a good system for precision testing of these models because the simple electronic structure containing a single valence electron makes calculations easier and the high atomic number leads to enhancement in some effects such as the parity non-conservation (PNC) interaction [3,6]. Recently, the $nD_{5/2}$ states of Cs have received attention [7–9] because of their promise as candidates for PNC measurements [10]. Apart from fundamental science, Cs is of immense importance because the definition of a second is based on the HFS in the $6S_{1/2}$ state of Cs.

In this article, we report the measurement of the HFS in the $7D_{5/2}$ state of Cs using Doppler-free two-photon spectroscopy in a cesium vapor cell. We determine the magnetic dipole coupling constant ($A$) and the electric quadrupole coupling constant ($B$) with precisions that represent at least 20-times improvement in precision over earlier reports [11–14]. We resolve the ambiguity in the sign of $B$ and establish it to be positive. Additionally, we determine the nuclear magnetic octupole coupling constant $C$, for which there are no prior reports. Further, given the possibility of a portable optical frequency standard at 767.2 nm based on the cesium $6S_{1/2} \to 7D_{5/2}$ Doppler-free two-photon transition in a vapor cell, we report the measurement of the ac Stark shift, collisional shift and pressure broadening of the cesium $6S_{1/2} \to 7D_{5/2}$ transition and provide valuable inputs for analysis of systematic effects in optical clocks.

The enhanced precision of our measurement is a result of two innovative schemes, one related to fluorescence detection and the other related to laser frequency tuning using an acousto-optic modulator (AOM). Like other experiments, we excite the $6S_{1/2} \to 7D_{5/2}$ two-photon transition using the Doppler-free geometry i.e. with counter propagating laser beams of wavelength 767.2 nm but, unlike previous reports [12–14], we choose to detect the direct fluorescence emitted by atoms while undergoing the $7D_{5/2} \to 6P_{3/2}$ transition (see Inset of Fig. 1). Our scheme results in higher signal-to-noise ratio (SNR) because the $7D_{5/2} \to 6P_{3/2}$ decay channel has the highest branching ratio (~76%) [15] and the emitted light at 698 nm cannot be reabsorbed by the Cs atoms. This may be contrasted with previous reports [12–14] detecting the $7P_{3/2} \to 6S_{1/2}$ cascade fluorescence at 456 nm that can be reabsorbed by the Cs atoms in the $6S_{1/2}$ state. The other improvement we achieve is due to the implementation of a technique for precise linear frequency scan of the laser using an AOM in cat's eye double-pass configuration. This solves the issue of non-linearly in frequency scans that has plagued many previous experiments.

A schematic diagram of our experimental setup is shown in Fig. 1. We use an external cavity diode laser (ECDL) combined with a Tapered Amplifier (TA) as the light source. The light is coupled out through an optical fiber to obtain a Gaussian beam with a $1/e^2$ diameter of ~1.55 mm. The light is then split into two beams – each of which is passed through their respective AOM and then send to their respective Cs vapor cell for Doppler-free two-photon spectroscopy. The first beam undergoes a double-pass through AOM 1, whose rf drive frequency is kept fixed at 106.5 MHz, before entering Cs cell 1 where it excites one of the $6S_{1/2} \to 7D_{5/2}$ hyperfine transitions. The fluorescence is collected, routed through a 700-nm Short-Pass (SP) filter, detected using an avalanche photodiode (APD) and provided as an input to the feedback loop for frequency stabilization of the laser – the laser is locked either to the $6S_{1/2}$ ($F = 3$) → $7D_{5/2}$ ($F' = 3$) or the $6S_{1/2}$ ($F = 4$) → $7D_{5/2}$ ($F' = 5$) transition depending on the experiment.

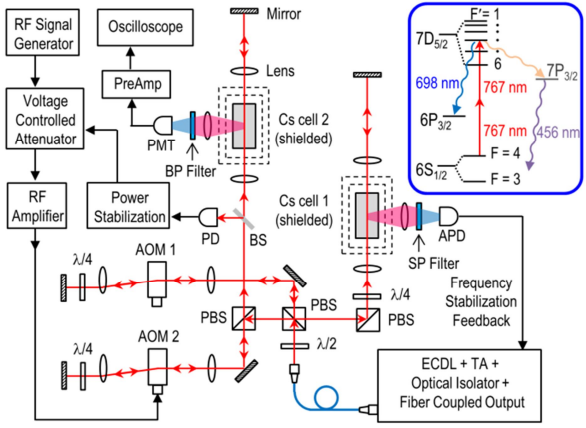

**Fig. 1.** Schematic representation of the experimental setup. Inset: the relevant energy level diagram of cesium (not to scale).

The second beam is send through AOM 2 in a carefully arranged double-pass cat's eye configuration [16] that eliminates the angular deflection of the beam when the rf frequency is tuned. The AOM 2 is driven by a rf signal generator (Stanford Research Systems SG386) with sub-Hz resolution whose frequency can scanned linearly in time using the internal ramp function. We scan the frequency of the rf source by 36 MHz leading to a laser frequency scan of 72 MHz. For a two photon transition this amount to a total scan of 144 MHz, sufficiently wide to cover the entire $7D_{5/2}$ state hyperfine spectrum spanning around 34 MHz. We scan the rf frequency at a rate of 1 Hz. In order to keep the laser beam power fixed as the AOM 2 frequency is tuned, a small fraction of the beam is picked off using a beam sampler (BS) and monitored on a photodiode (PD). The PD voltage is compared with a reference voltage to generate a feedback signal that is sent to the voltage controlled attenuator (VCA) to control the rf power to AOM 2 and keep the laser power constant (within 0.2%) at the input of Cs cell 2. The beam is then weakly focused ($1/e^2$ radius $r = 63\pm2$ μm) into the spectroscopy cell (Cs cell 2), enclosed in 2-layers of mu-metal shielding that reduces the stray magnetic field to < 2 mG. The linearly polarized laser beam excites the $6S_{1/2}$ → $7D_{5/2}$ two-photon transition and the fluorescence is collected, routed through a 10-nm Band-Pass (BP) filter centered at 695 nm and detected on a photo-multiplier tube (PMT). The PMT signal is fed to a low noise pre-amplifier and then sent to a digital storage oscilloscope (DSO) where the data is stored. We store single scan data as well as data averaged over 4 scans and 16 scans on the DSO.

Figures 2(a) and 2(b) show the Doppler-free two-photon spectra obtained when the laser is scanned across the $6S_{1/2}$ ($F = 3$) → $7D_{5/2}$ ($F' = 1 – 5$) and the $6S_{1/2}$ ($F = 4$) → $7D_{5/2}$ ($F' = 2 – 6$) transitions, respectively. All transitions obeying the $\Delta F = 0, \pm1, \pm2$ selection rules are observed. Also shown are the multi-peak fits to a combination of five Voigt profiles. The high SNR results in excellent fits and better resolution compared to prior reports. Notably, the $F' = 1$ and $F' = 2$ levels are partly resolved which was not the case in earlier reports [12–14]. We determine the line center and the linewidth from the fits, along with the peak height and the overall offset (which turns out to be small).

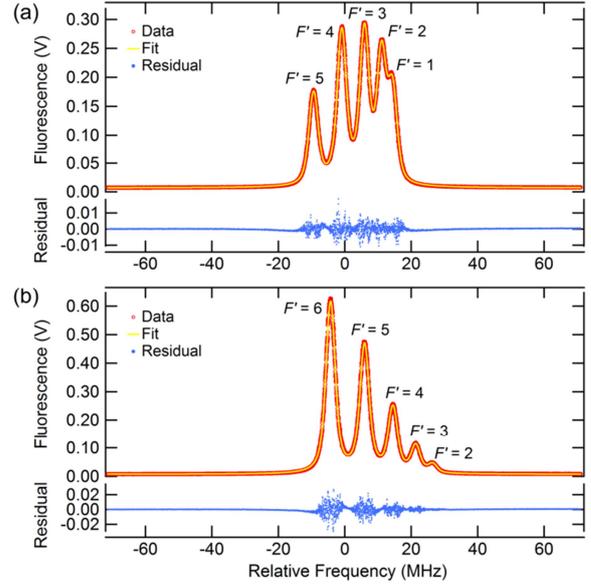

**Fig. 2.** Doppler-free two-photon spectra of Cs for (a) the $6S_{1/2}$ ($F = 3$) → $7D_{5/2}$ ($F'$) and (b) the $6S_{1/2}$ ($F = 4$) → $7D_{5/2}$ ($F'$) transitions.

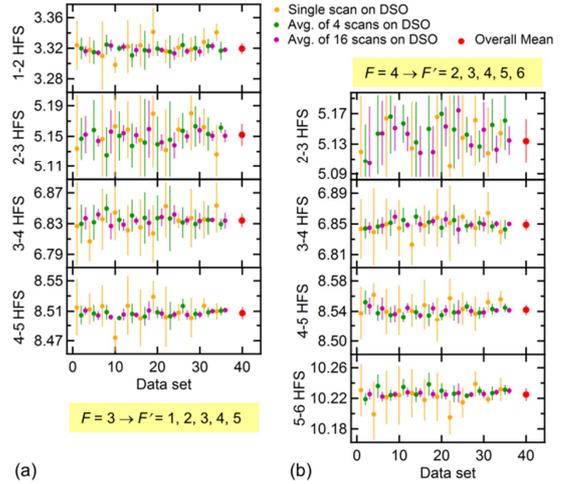

**Fig. 3.** Repeatability of hyperfine splitting measurements conducted over several weeks. (a) $6S_{1/2}$ ($F = 3$) → $7D_{5/2}$ ($F'$) transitions. (b) $6S_{1/2}$ ($F = 4$) → $7D_{5/2}$ ($F'$) transitions. Laser power is 231 mW and Cs cell 2 is at ∼90°C. The last data point (red) is the overall mean of all data points. All other data points are average of seven measurements.

In Figures 3(a) and 3(b), we plot the HFS measured on different days to demonstrate the repeatability of the data. These experiments were done with Cs cell 2 at ∼90°C and a laser power ($P$) of 231 mW at the input of the cell. The HFS measured for excitation from either of the two ground states, $6S_{1/2}$ ($F = 3$) or $6S_{1/2}$ ($F = 4$), are similar, thus providing a consistency check. The measured HFS are reported in Table 1 along with values from previous reports for comparison. We determine the HFS with a precision of ∼10 kHz, representing at least an order of magnitude improvement over earlier reports [12–14].

**Table 1. Hyperfine splitting (in MHz) between consecutive $F'$ levels measured in this work, compared to earlier reports.**

| HFS | This work $6S_{1/2}$ $F=3$ | This work $6S_{1/2}$ $F=4$ | Ref. [12] [a] | Ref. [13] | Ref. [14] [b] |
|---|---|---|---|---|---|
| 1 – 2 | 3.320 (0.007) | | 3.38 (0.15) | | |
| 2 – 3 | 5.152 (0.015) | 5.134 (0.029) | 5.09 (0.17) | 5.90 (0.25) | 5.82 (0.25) |
| 3 – 4 | 6.834 (0.009) | 6.849 (0.007) | 6.83 (0.13) | 7.46 (0.25) | 7.40 (0.25) |
| 4 – 5 | 8.508 (0.008) | 8.542 (0.007) | 8.59 (0.08) | 9.04 (0.35) | 9.01 (0.25) |
| 5 – 6 | | 10.225 (0.009) | 10.39 (0.28) | 10.39 (0.35) | 10.20 (0.25) |

[a] HFS and errors are calculated from the reported value of $A$ and $B$.
[b] The errors were estimated to be ∼0.25 MHz.

In Figures 4(a) and 4(b) we plot the fluorescence signal as a function of the square of the laser power ($P^2$) demonstrating the expected quadratic dependence of the signal on $P$. Figures 4(c) and 4(d) show the shift in the peak position as a function of the laser power. The peak positions of all the $F'$ levels shift linearly with $P$ and at the same rate implying that the HFS remains unchanged with variations of $P$ i.e. systematic effects on HFS due to light shift is negligible (see Supplement 1). From the slope of these plots, we determine the ac Stark shift to be -46±4 Hz/(W/cm²), where we have considered 86% transmission through each window of the Cs cell leading to a total bidirectional power of $1.5P$ in the interaction region and the peak intensity $I = 1.5P/(\pi r^2/2)$. This light shift must be taken into account in any frequency standard based on the $6S_{1/2} \to 7D_{5/2}$ two-photon transition. We do not observe any appreciable power broadening of the lines.

In Figures 5(a) and 5(b) we plot the shift in the peak positions as a function of the Cs vapor pressure ($p$), which is varied by changing the temperature of the Cs cell 2 in the range 50°C to 159°C. The peak positions of all the $F'$ levels shift linearly with $p$ but at the same rate (-24±2 kHz/mTorr and -36±3 kHz/mTorr for the $F = 3 \to F'$ and $F = 4 \to F'$ transitions, respectively) implying that the HFS remains unchanged with variations of the temperature which is also evident from the overlapping data points for all $F'$ levels. We performed experiments with two different Cs cells and found the HFS to be independent of the cell used. Figures 5(c) And 5(d) show the variation of the linewidth (the Lorentzian part of the Voigt fit) as the Cs vapor pressure is changed. The linewidths increase linearly with vapor pressure. While the linewidths of different $F'$ level appear to be slightly different, this may be an artifact due to fitting of closely spaced levels. We therefore consider the average value. On an average, the linewidth increases linearly with vapor pressure with a slope 101±19 kHz/mTorr and zero-pressure-intercept of 2.1±0.3 MHz. The measured linewidth is slightly higher than the expected linewidth (∼1.7 MHz) [15,17] because of contributions from transit time broadening and the laser linewidth.

We repeated the experiments with a residual magnetic field of 15-20 mG and did not observe any change in the HFS compared to experiments done with stray fields below 2 mG. This implies that the Zeeman effect does not play a role in the measurement of HFS and this result is expected for linearly polarized light used in the experiment. This is good news for applications in portable frequency standards. We performed experiments without the focusing lens and did not observe any change in the measured HFS.

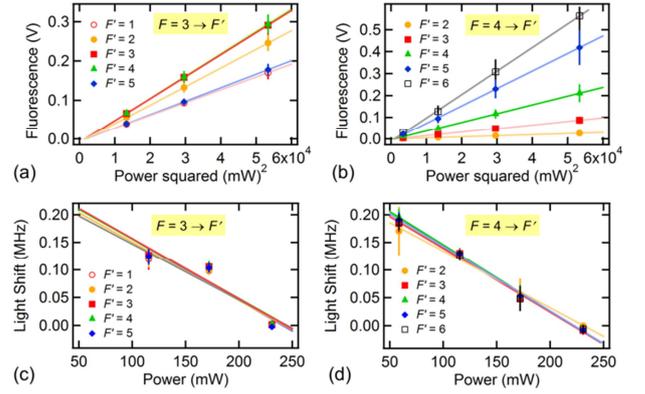

**Fig. 4.** (a), (b) The fitted peak height plotted against the square of the laser power $P$, along with fits to a quadratic expression in $P$. (c), (d) The shift in the line position (relative to data taken at 231 mW) plotted against the laser power $P$, along with fits to a linear expression in $P$.

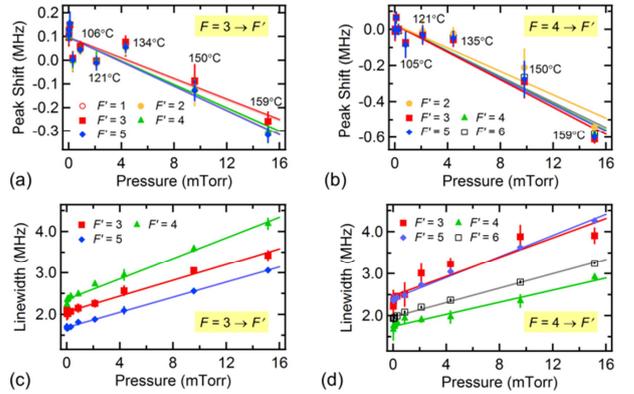

**Fig. 5.** (a), (b) The shift in the peak position (relative to data taken at 90°C) plotted against the vapor pressure in Cs cell 2, along with fits to a linear expression in $p$. (c), (d) The linewidths of the well-resolved lines plotted against the vapor pressure in Cs cell 2, along with fits to a linear expression in $p$.

With all the tangible systemic effects measured and found to be negligible in the determination of HFS at our level of precision, we can focus on determining the hyperfine coupling constants $A$, $B$ and $C$. The hyperfine interaction energy for a hyperfine level $F'$ in the $7D_{5/2}$ state is given by [18]:

$$E(F') = A\frac{K}{2} + B\frac{3K(K+1) - 2205/4}{1680}$$
$$+ C\frac{5K^2(K/4+1) - 6175K/16 - 11025/16}{1575/2} \qquad (1)$$

**Table 2. Hyperfine coupling constants measured in this work and a comparison with prior reports.**

|  | This work | Ref. [12] | Ref. [13] | Ref. [14] | Ref. [11] Theory |
|---|---|---|---|---|---|
| A (in MHz) | -1.70867 (0.00062) | -1.717 (0.015) | -1.81 (0.05) | -1.79 (0.05) | -1.42 |
| B (in MHz) | 0.050 (0.014) | -0.18 (0.52) | 1.01 (1.06) | 1.05 (0.29) | — |
| C (in kHz) | 0.4 (1.4) | — | — | — | — |

where $K = F'(F'+1) - 49/2$. This equation is used to find the expression for HFS, $E(F'_a) - E(F'_b)$, between any two hyperfine levels $F'_a$ and $F'_b$, leading to a set of expressions which are equated to the respective measured HFS (see Supplement 1). We then perform a global fit to all the equations to obtain the values of $A$, $B$ and $C$. We find that the HFS data measured for excitation from the $6S_{1/2}$ ($F = 4$) state (Column 3 of Table 1) gives significantly lower fitting errors compared the data for excitation from the $6S_{1/2}$ ($F = 3$) state. This happens because the 1–2 HFS is not well resolved for excitation from the $6S_{1/2}$ ($F = 3$) state. Further, we find that for excitation from the $6S_{1/2}$ ($F = 4$) state, there is a clear progression in the determined values of $A$, $B$ and $C$ depending on whether or not the last term in Eq. (1) is included. On excluding the last term in Eq. (1), we get $A$ = -1.70881 ± 0.00040 MHz and $B$ = 0.05347 ± 0.01033 MHz, both of which change only slightly on including the last term, leading to $A$ = -1.70867 ± 0.00062 MHz, $B$ = 0.05049 ± 0.01437 MHz and $C$ = 0.00042 ± 0.00141 MHz. The numbers after the ± sign represent the global fit errors. We recommend the values of $A$, $B$ and $C$ as determined from the data obtained from excitation from the $6S_{1/2}$ ($F = 4$) state. The values are reported in Table 2. The values of $A$, $B$ and $C$ obtained on using the average of the measured HFS (i.e. average of columns 2 and 3) are consistent with these values if the $F'$ =1 level is excluded from the global fit (see Supplement 1). A comparison with earlier results (see Table 2) shows that our technique offers at least a 20-fold improvement in the precision of $A$ and $B$, and also yields the value of $C$ which was so far undetermined. The available theoretical calculation of the hyperfine coupling constant $A$ = -1.42 MHz [11] does not agree well with any of the experimental reports.

We note that further improvement in the determination of $C$ can be obtained using the values of HFS reported in column 3 of Table 1 and incorporating the second order corrections in HFS from high-precision theoretical calculations, such as in Refs. [19,20], which are not yet available for the $7D_{5/2}$ state. Such theoretical calculations will also provide the value of $C/\Omega$, from which the nuclear octupole moment $\Omega$ of $^{133}$Cs can be determined using the known value of $C$. This is of interest because there is, so far, only one experimental determination of $\Omega$ for $^{133}$Cs using hyperfine spectroscopy which deviates from the nuclear shell model prediction by a factor of 40 [21].

In conclusion, we demonstrate a technique for high resolution measurement of HFS based on scanning the laser frequency using an AOM driven by a stable high resolution rf source. The scheme can be applied to HFS measurement of atoms, molecules and ions, both in vapor cells and in trapped samples at ultracold temperatures. We use the scheme to determine the HFS of Cs atoms with unprecedented precision using Doppler-free two-photon spectroscopy in a Cs vapor cell. The technique also allowed the determination of light shift and collisional shift, both of which are important for making portable frequency standard based on the Cs $6S_{1/2} \to 7D_{5/2}$ two-photon transition at 767 nm.

**Funding.** Department of Atomic Energy, Government of India, Project Identification No. RTI4002.

**Acknowledgment**. We thank Bijaya K. Sahoo for discussions initiated during the ICTS Meeting on Trapped Atoms, Molecules and Ions, 2022 (code: ICTS/TAMIONs-2022/5).

**Disclosures.** The authors declare no conflicts of interest.

**Data availability.** Data underlying the results presented in this paper are not publicly available at this time but may be obtained from the authors upon reasonable request.

**Supplemental document.** See Supplement 1 for supporting content.

# Supplementary Information:

## 1. Hyperfine splitting (HFS) measured at different powers:

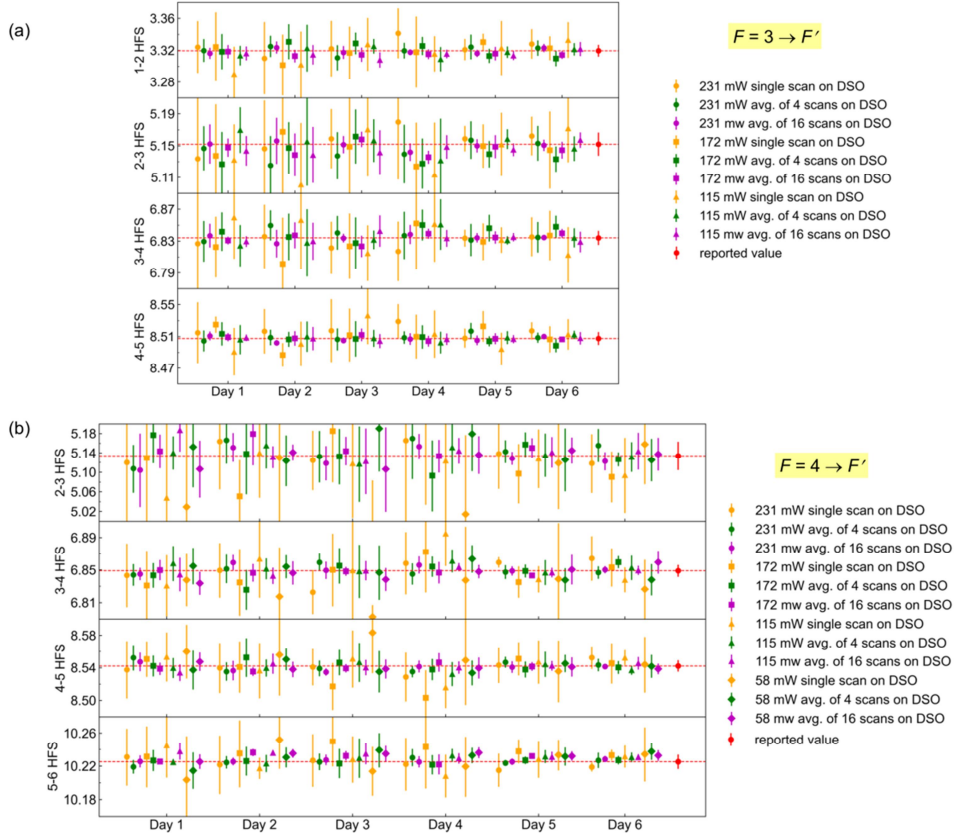

**Fig. S1.** The measured HFS of the 7D$_{5/2}$ state at different laser powers and different days for excitation from (a) the 6S$_{1/2}$ ($F$ = 3) state and (b) the 6S$_{1/2}$ ($F$ = 4) state.

## 2. Determination of hyperfine coupling constants:

Expressions for hyperfine splitting (HFS), $E(F'_a) - E(F'_b)$, between any two hyperfine levels $F'_a$ and $F'_b$:

$$E(F' = 6) - E(F' = 5) = 6A + \frac{18}{35}B + \frac{144}{35}C \quad \text{(S1)}$$

$$E(F' = 6) - E(F' = 4) = 11A + \frac{11}{20}B - \frac{11}{35}C \quad \text{(S2)}$$

$$E(F' = 6) - E(F' = 3) = 15A + \frac{9}{28}B - \frac{15}{7}C \quad \text{(S3)}$$

$$E(F' = 6) - E(F' = 2) = 18A + \frac{24}{35}C \quad \text{(S4)}$$

$$E(F' = 5) - E(F' = 4) = 5A + \frac{1}{28}B - \frac{31}{7}C \quad \text{(S5)}$$

$$E(F' = 5) - E(F' = 3) = 9A - \frac{27}{140}B - \frac{219}{35}C \quad \text{(S6)}$$

$$E(F' = 5) - E(F' = 2) = 12A - \frac{18}{35}B - \frac{24}{7}C \quad \text{(S7)}$$

$$E(F' = 4) - E(F' = 3) = 4A - \frac{8}{35}B - \frac{64}{35}C \quad \text{(S8)}$$

$$E(F' = 4) - E(F' = 2) = 7A - \frac{11}{20}B + C \quad \text{(S9)}$$

$$E(F' = 3) - E(F' = 2) = 3A - \frac{9}{28}B + \frac{99}{35}C \quad \text{(S10)}$$

$$E(F' = 5) - E(F' = 1) = 14A - \frac{4}{5}B + \frac{8}{5}C \quad \text{(S11)}$$

$$E(F' = 4) - E(F' = 1) = 9A - \frac{117}{140}B + \frac{211}{35}C \quad \text{(S12)}$$

$$E(F' = 3) - E(F' = 1) = 5A - \frac{17}{28}B + \frac{55}{7}C \quad \text{(S13)}$$

$$E(F' = 2) - E(F' = 1) = 2A - \frac{2}{7}B + \frac{176}{35}C \quad \text{(S14)}$$

Equations S1–S10 are used in case of excitation from the $6S_{1/2}$ ($F$ = 4) state. Equations S5–S14 are used in case of excitation from the $6S_{1/2}$ ($F$ = 3) state.

**Table S1:** Comparison of values $A$, $B$ and $C$, obtained by global fitting to the equations above under different conditions e.g. using the HFS measured for excitation from $6S_{1/2}$ ($F$ = 4) or the $6S_{1/2}$ ($F$ = 3) state or their average, setting $C$ = 0, keeping $C$ as a free parameter, excluding the $F'$ = 1 level from the fit etc. We find that excluding the $F'$ = 1 level, which is not fully resolved, results in better global fits.

| Values of $A$, $B$ and $C$ (in MHz) obtained from the global fit | $F'$ levels | Calculated HFS (in MHz) determined using $A$, $B$ and $C$ | Measured HFS (in MHz) for excitation from $F$ = 4 state | Difference (in kHz): Calculated – Measured from $F$ = 4 state | Measured HFS (in MHz) for excitation from $F$ = 3 state | Difference (in kHz): Calculated – Measured from $F$ = 3 state |
|---|---|---|---|---|---|---|
| **$A$ = -1.70881(40)** **$B$ = 0.05347(1033)** **$C$ = 0 (fixed)** $A$ and $B$ obtained by fitting equations S1-S10 to HFS measured for excitation from the $6S_{1/2}$ ($F$ = 4) state. | 6 – 5 | -10.22536 | -10.225 | 0 | | |
| | 6 – 4 | -18.76750 | -18.767 | -1 | | |
| | 6 – 3 | -25.61496 | -25.616 | 1 | | |
| | 6 – 2 | -30.75858 | -30.75 | -9 | | |
| | 5 – 4 | -8.54214 | -8.542 | 0 | -8.508 | -34 |
| | 5 – 3 | -15.38960 | -15.391 | 1 | -15.342 | -48 |
| | 5 – 2 | -20.53322 | -20.525 | -8 | -20.494 | -39 |
| | 4 – 3 | -6.84746 | -6.849 | 2 | -6.834 | -13 |
| | 4 – 2 | -11.99108 | -11.983 | -8 | -11.986 | -5 |
| | 3 – 2 | -5.14362 | -5.134 | -10 | -5.152 | 8 |
| | 5 – 1 | -23.96612 | | | -23.813 | -153 |
| | 4 – 1 | -15.42398 | | | -15.306 | -118 |
| | 3 – 1 | -8.57651 | | | -8.472 | -105 |
| | 2 – 1 | -3.43290 | | | -3.320 | -113 |
| **$A$ = -1.70867(62)** **$B$ = 0.05049(1437)** **$C$ = 0.00042(141)** $A$ and $B$ obtained by fitting equations S1-S10 to HFS measured for excitation from the $6S_{1/2}$ ($F$ = 4) state. *notice that A, B changed only slightly compared to the previous case. | 6 – 5 | -10.22433 | -10.225 | 1 | | |
| | 6 – 4 | -18.76773 | -18.767 | -1 | | |
| | 6 – 3 | -25.61472 | -25.616 | 1 | | |
| | 6 – 2 | -30.75577 | -30.75 | -6 | | |
| | 5 – 4 | -8.54341 | -8.542 | -1 | -8.508 | -35 |
| | 5 – 3 | -15.39040 | -15.391 | 1 | -15.342 | -48 |
| | 5 – 2 | -20.53145 | -20.525 | -6 | -20.494 | -37 |
| | 4 – 3 | -6.84699 | -6.849 | 2 | -6.834 | -13 |
| | 4 – 2 | -11.98804 | -11.983 | -5 | -11.986 | -2 |
| | 3 – 2 | -5.14105 | -5.134 | -7 | -5.152 | 11 |
| | 5 – 1 | -23.96110 | | | -23.813 | -148 |
| | 4 – 1 | -15.41769 | | | -15.306 | -112 |
| | 3 – 1 | -8.57070 | | | -8.472 | -99 |
| | 2 – 1 | -3.42965 | | | -3.320 | -110 |
| **$A$ = -1.70753(86)** **$B$ = -0.0682 (142)** **$C$ = 0 (fixed)** $A$ and $B$ obtained by fitting equations S5-S14 to HFS measured for excitation from the $6S_{1/2}$ ($F$ = 3) state. | 6 – 5 | -10.28025 | -10.225 | -55 | | |
| | 6 – 4 | -18.82034 | -18.767 | -53 | | |
| | 6 – 3 | -25.63487 | -25.616 | -19 | | |
| | 6 – 2 | -30.73554 | -30.75 | 14 | | |
| | 5 – 4 | -8.54009 | -8.542 | 2 | -8.508 | -32 |
| | 5 – 3 | -15.35462 | -15.391 | 36 | -15.342 | -13 |
| | 5 – 2 | -20.45529 | -20.525 | 70 | -20.494 | 39 |
| | 4 – 3 | -6.81453 | -6.849 | 34 | -6.834 | 19 |
| | 4 – 2 | -11.91520 | -11.983 | 68 | -11.986 | 71 |
| | 3 – 2 | -5.10067 | -5.134 | 33 | -5.152 | 51 |
| | 5 – 1 | -23.85086 | | | -23.813 | -38 |
| | 4 – 1 | -15.31077 | | | -15.306 | -5 |
| | 3 – 1 | -8.49624 | | | -8.472 | -24 |
| | 2 – 1 | -3.39557 | | | -3.320 | -76 |

| | | | | | |
|---|---|---|---|---|---|
| $A$ = -1.67791(248)<br>$B$ = 0.4563 (436)<br>$C$ = 0.0281(22)<br><br>$A$ and $B$ obtained by fitting equations S5-S14 to HFS measured for excitation from the $6S_{1/2}$ ($F$ = 3) state.<br><br>*notice that A, B have changed significantly compared to the previous case.* | 6 – 5 | -9.71718 | -10.225 | 508 | | |
| | 6 – 4 | -18.21488 | -18.767 | 552 | | |
| | 6 – 3 | -25.08220 | -25.616 | 534 | | |
| | 6 – 2 | -30.18311 | -30.75 | 567 | | |
| | 5 – 4 | -8.49770 | -8.542 | 44 | -8.508 | 10 |
| | 5 – 3 | -15.36502 | -15.391 | 26 | -15.342 | -23 |
| | 5 – 2 | -20.46593 | -20.525 | 59 | -20.494 | 28 |
| | 4 – 3 | -6.86732 | -6.849 | -18 | -6.834 | -33 |
| | 4 – 2 | -11.96824 | -11.983 | 15 | -11.986 | 18 |
| | 3 – 2 | -5.10092 | -5.134 | 33 | -5.152 | 51 |
| | 5 – 1 | -23.81082 | | | -23.813 | 2 |
| | 4 – 1 | -15.31312 | | | -15.306 | -7 |
| | 3 – 1 | -8.44580 | | | -8.472 | 26 |
| | 2 – 1 | -3.34489 | | | -3.320 | -25 |
| $A$ = -1.70215(98)<br>$B$ = 0.1281(215)<br>$C$ = 0 (fixed)<br><br>$A$ and $B$ obtained by fitting equations S5-S10 to HFS measured for excitation from the $6S_{1/2}$ ($F$ = 3) state.<br>*Excludes all F' = 1 levels from the fit.* | 6 – 5 | -10.14702 | -10.225 | 78 | | |
| | 6 – 4 | -18.65320 | -18.767 | 114 | | |
| | 6 – 3 | -25.49108 | -25.616 | 125 | | |
| | 6 – 2 | -30.63870 | -30.75 | 111 | | |
| | 5 – 4 | -8.50618 | -8.542 | 36 | -8.508 | 2 |
| | 5 – 3 | -15.34406 | -15.391 | 47 | -15.342 | -2 |
| | 5 – 2 | -20.49168 | -20.525 | 33 | -20.494 | 2 |
| | 4 – 3 | -6.83788 | -6.849 | 11 | -6.834 | -4 |
| | 4 – 2 | -11.98551 | -11.983 | -3 | -11.986 | 0 |
| | 3 – 2 | -5.14763 | -5.134 | -14 | -5.152 | 4 |
| | 5 – 1 | -23.93258 | | | -23.813 | -120 |
| | 4 – 1 | -15.42641 | | | -15.306 | -120 |
| | 3 – 1 | -8.58853 | | | -8.472 | -117 |
| | 2 – 1 | -3.44090 | | | -3.320 | -121 |
| $A$ = -1.70509 (467)<br>$B$ = 0.08567(6934)<br>$C$ = -0.00325 (505)<br><br>$A$ and $B$ obtained by fitting equations S5-S10 to HFS measured for excitation from the $6S_{1/2}$ ($F$ = 3) state.<br>*Excludes all F' = 1 levels from the fit.*<br><br>*notice that A, B have changed significantly compared to previous case.* | 6 – 5 | -10.19985 | -10.225 | 25 | | |
| | 6 – 4 | -18.70785 | -18.767 | 59 | | |
| | 6 – 3 | -25.54185 | -25.616 | 74 | | |
| | 6 – 2 | -30.69385 | -30.75 | 56 | | |
| | 5 – 4 | -8.50800 | -8.542 | 34 | -8.508 | 0 |
| | 5 – 3 | -15.34200 | -15.391 | 49 | -15.342 | 0 |
| | 5 – 2 | -20.49400 | -20.525 | 31 | -20.494 | 0 |
| | 4 – 3 | -6.83400 | -6.849 | 15 | -6.834 | 0 |
| | 4 – 2 | -11.98600 | -11.983 | -3 | -11.986 | 0 |
| | 3 – 2 | -5.15200 | -5.134 | -18 | -5.152 | 0 |
| | 5 – 1 | -23.94500 | | | -23.813 | -132 |
| | 4 – 1 | -15.43700 | | | -15.306 | -131 |
| | 3 – 1 | -8.60300 | | | -8.472 | -131 |
| | 2 – 1 | -3.45100 | | | -3.320 | -131 |
| $A$ = -1.70842(27)<br>$B$ = 0.0461(38)<br>$C$ = 0 (fixed)<br><br>$A$ and $B$ obtained by fitting equations S1-S10 to **average** of HFS measured for excitation from the $6S_{1/2}$ ($F$ = 3) and $6S_{1/2}$ ($F$ = 4) states.<br>*Excludes all F' = 1 levels from the fit.* | 6 – 5 | -10.22681 | -10.225 | -2 | | |
| | 6 – 4 | -18.76727 | -18.767 | 0 | | |
| | 6 – 3 | -25.61148 | -25.616 | 5 | | |
| | 6 – 2 | -30.75156 | -30.75 | -2 | | |
| | 5 – 4 | -8.54045 | -8.542 | 2 | -8.508 | -32 |
| | 5 – 3 | -15.38467 | -15.391 | 6 | -15.342 | -43 |
| | 5 – 2 | -20.52475 | -20.525 | 0 | -20.494 | -31 |
| | 4 – 3 | -6.84422 | -6.849 | 5 | -6.834 | -10 |
| | 4 – 2 | -11.98430 | -11.983 | -1 | -11.986 | 2 |
| | 3 – 2 | -5.14008 | -5.134 | -6 | -5.152 | 12 |
| | 5 – 1 | -23.95476 | | | -23.813 | -142 |
| | 4 – 1 | -15.41431 | | | -15.306 | -108 |
| | 3 – 1 | -8.57009 | | | -8.472 | -98 |
| | 2 – 1 | -3.43001 | | | -3.320 | -110 |
| $A$ = -1.70841(27)<br>$B$ = 0.0448(41)<br>$C$ = -0.00094(118)<br><br>$A$ and $B$ obtained by fitting equations S1-S10 to **average** of HFS | 6 – 5 | -10.23129 | -10.225 | -6 | | |
| | 6 – 4 | -18.76757 | -18.767 | -1 | | |
| | 6 – 3 | -25.60974 | -25.616 | 6 | | |
| | 6 – 2 | -30.75202 | -30.75 | -2 | | |
| | 5 – 4 | -8.53629 | -8.542 | 6 | -8.508 | -28 |
| | 5 – 3 | -15.37845 | -15.391 | 13 | -15.342 | -36 |
| | 5 – 2 | -20.52074 | -20.525 | 4 | -20.494 | -27 |
| measured for excitation from the $6S_{1/2}$ ($F = 3$) and $6S_{1/2}$ ($F = 4$) states. *Excludes all F' = 1 levels from the fit.* *values comparable to excitation from the $6S_{1/2}$ (F = 4) state.* | 4 – 3 | -6.84216 | -6.849 | 7 | -6.834 | -8 |
| | 4 – 2 | -11.98445 | -11.983 | -1 | -11.986 | 2 |
| | 3 – 2 | -5.14229 | -5.134 | -8 | -5.152 | 10 |
| | 5 – 1 | -23.95508 | | | -23.813 | -142 |
| | 4 – 1 | -15.41880 | | | -15.306 | -113 |
| | 3 – 1 | -8.57664 | | | -8.472 | -105 |
| | 2 – 1 | -3.43435 | | | -3.320 | -114 |